\documentclass[conference]{IEEEtran}
\IEEEoverridecommandlockouts
\usepackage{hyperref}
\usepackage{cite}
\usepackage{amsmath,amssymb,amsfonts}
\usepackage{algorithmic}
\usepackage{algorithm}
\usepackage{graphicx}
\usepackage{caption}
\usepackage{textcomp}
\usepackage{booktabs}
\usepackage{xcolor}
\usepackage{subcaption}

\def\BibTeX{{\rm B\kern-.05em{\sc i\kern-.025em b}\kern-.08em
    T\kern-.1667em\lower.7ex\hbox{E}\kern-.125emX}}
\begin{document}

\title{Collaborative Inference Acceleration with Non-Penetrative Tensor Partitioning\\

}


\author{

    \IEEEauthorblockN{Zhibang Liu, Chaonong Xu$^{\ast}$\thanks{$^{\ast}$Corresponding author}, Zhenjie Lv, Zhizhuo Liu and Suyu Zhao}
    \IEEEauthorblockA{Beijing Key Lab of Petroleum Data Mining, 
 China University of Petroleum, Beijing, China\\
    2021310705@student.cup.edu.cn, xuchaonong@cup.edu.cn
    }
    
    
}


\maketitle

\begin{abstract}
The inference of large-sized images on Internet of Things (IoT) devices is commonly hindered by limited resources, while there are often stringent latency requirements for Deep Neural Network (DNN) inference. Currently, this problem is generally addressed by collaborative inference, where the large-sized image is partitioned into multiple tiles, and each tile is assigned to an IoT device for processing. However, since significant latency will be incurred due to the communication overhead caused by tile sharing, the existing collaborative inference strategy is inefficient for convolutional computation, which is indispensable for any DNN. To reduce it, we propose Non-Penetrative Tensor Partitioning (NPTP), a fine-grained tensor partitioning method that reduces the communication latency by minimizing the communication load of tiles shared, thereby reducing inference latency. We evaluate NPTP with four widely-adopted DNN models. Experimental results demonstrate that NPTP achieves a 1.44-1.68× inference speedup relative to CoEdge, a state-of-the-art (SOTA) collaborative inference algorithm.
\end{abstract}


\section{Introduction}
Recent years have witnessed the growing prevalence of the Internet of Things (IoT) and the thriving of Deep Neural Networks (DNN), which have facilitated the deployment and inference of intelligent models on edge devices, presenting advantages for a variety of use cases\cite{cheng2023ai,ren2023survey,qu2023stochastic}. This trend also drives innovation in fields such as smart medicine\cite{zhuang2023arrhythmia}, smart factories\cite{soori2023internet}, and autonomous driving\cite{chib2023recent}, where the Artificial Intelligence of Things (AIoT) provides more intelligent and convenient services. 

However, the deployment and inference of DNN models on IoT devices currently face challenges from three aspects: application requirements, model size, and device attributes. First, for user experience and production safety considerations, such as heart rate monitoring\cite{zhafira2024implementation} and machine fault monitoring\cite{zhang2015iot}, the delay in model inference has a high demand for real-time performance, usually requiring milliseconds to complete the calculation and make the corresponding warning. Second, as the model size increases, the memory footprint and computational complexity for deploying intelligent models increase dramatically. Third, individual IoT devices tend to have limited storage and computing power due to the cost and size constraints\cite{lin2020mcunet,lin2021mcunetv2,sadiq2023enabling}.

\begin{figure}[!t]
\centering
\includegraphics[width=3in]{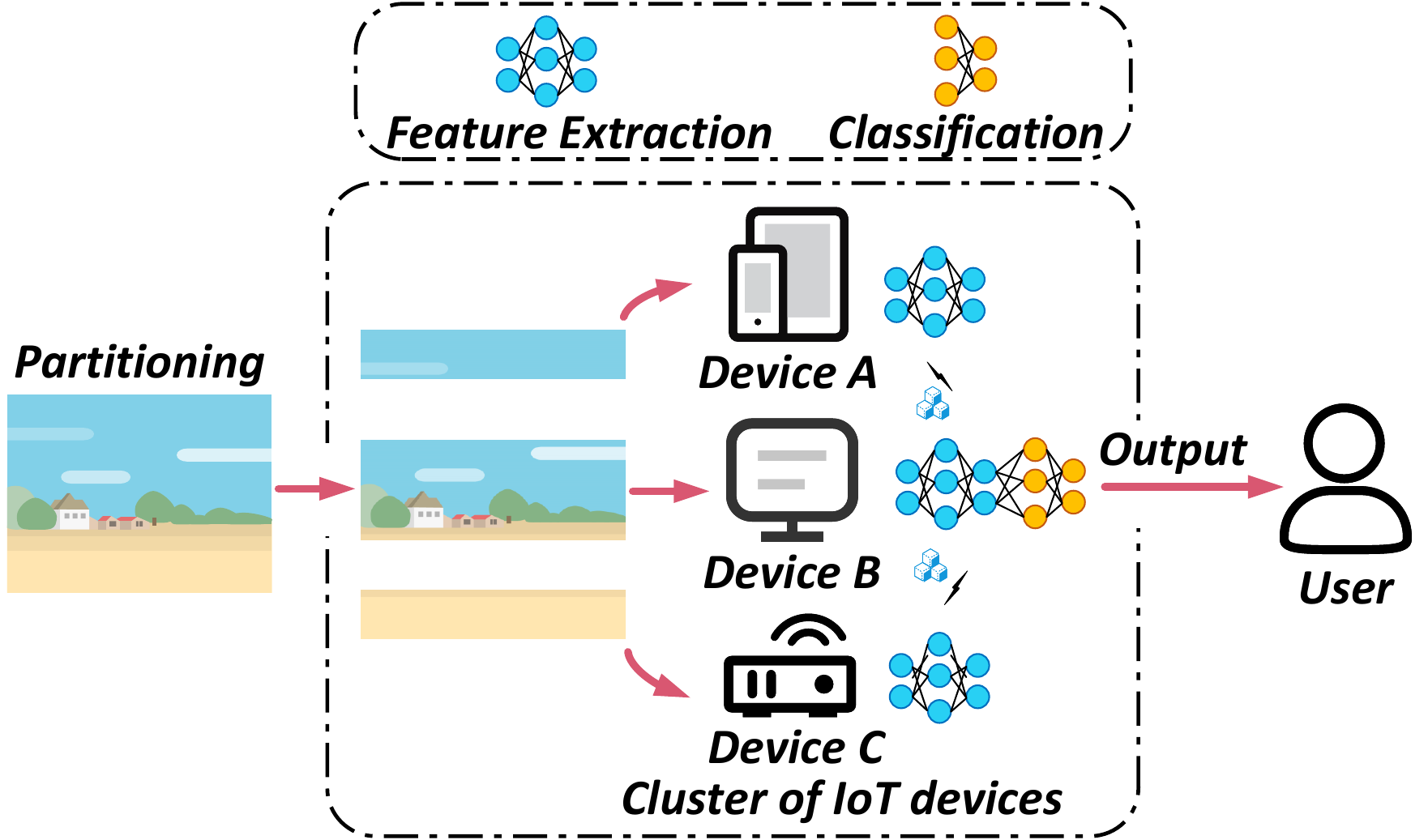}
\caption{An overview of collaborative image inference in an IoT scenario.}
\label{fig_1}
\end{figure}

In order to solve the problem, a promising approach is to exploit multiple IoT devices for collaborative inference\cite{shlezinger2022collaborative,wu2020accuracy,rodriguez2023horizontally,huang2022toward}. Widely adopted techniques for dealing with the above challenges in collaborative inference include data parallelism\cite{li2020pytorch}, model parallelism \cite{zhuang2023optimizing}, and hybrid parallelism \cite{liao2023accelerating}. Taking the image classification application in Fig. 1 as an example, the feature extraction part in the original model is replicated and deployed separately on devices A, B, and C. The input image is partitioned into three parts and fed to devices to generate three feature maps. In the classification phase, the three feature maps are aggregated at some device (B in this figure) to complete the remaining classification task.

However, since the convolution operations in the feature extraction layers of a DNN model are computed using a sliding window, the penetrative partition of the original image, as shown in Fig. 1, can lead to incomplete input data for the convolution process on one device. As a result, it has to retrieve the missing boundary image data (or sharing data in another name) from neighboring devices, which can incur significant inter-device communication overheads. 

Fortunately, we found that with a non-penetrative partitioning method, the inter-device communication overhead incurred from retrieving missing boundary data can be significantly reduced. Fig. 2 illustrates a typical convolution operation under both the penetrative and non-penetrative schemes. Fig. 2(a) reveals the case of penetrative partition, and the red and purple boxes indicate two different positions where the convolution kernel slides over the feature map. Here, we assume a standard convolution kernel size of 3×3 with a stride of 1. Device A needs to obtain the data from the 3rd row of the feature map from device B for the calculation at sliding window position 1. Similarly, for the calculation at sliding window position 2, device B needs to obtain the data from the 5th row of the feature map located on device C. The total amount of shared data is 24 units. In Fig. 2(b), where the non-penetrative partitioning scheme is applied, the amount of shared data decreases to 18 units. This reduces the inter-device communication overhead by 25\%.

\begin{figure}[!t]
\centering
\includegraphics[width=3.5in]{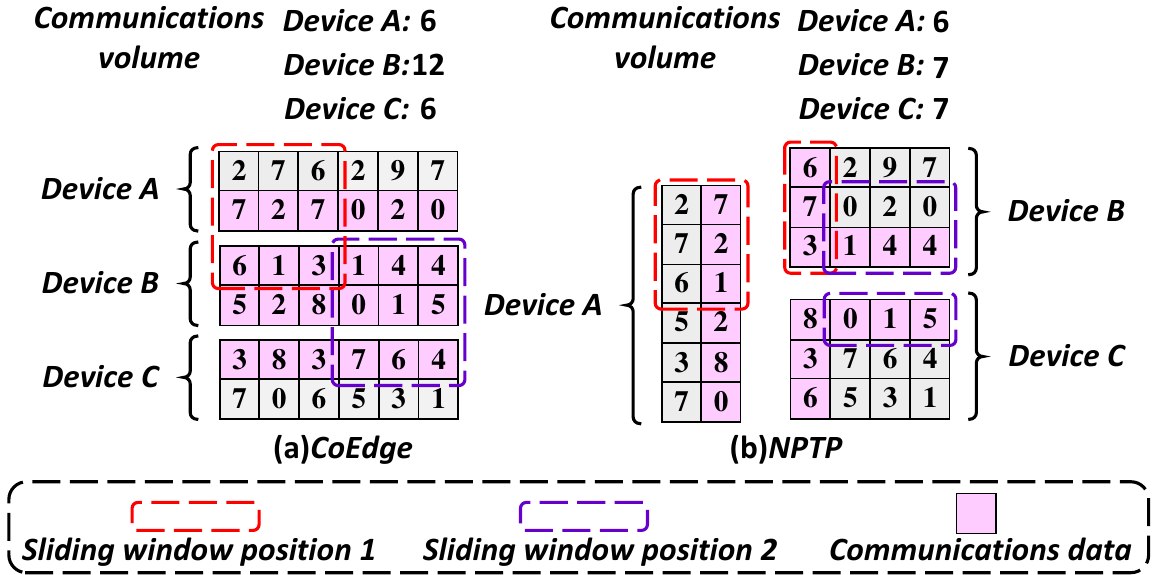}
\caption{An example of penetrative and non-penetrative image partitioning approaches for collaborative inference across three devices.}
\end{figure}

Based on the aforementioned observation, we propose Non-Penetrative Tensor Partitioning (NPTP), a distributed collaborative inference approach for AIoT devices. To obtain the optimal tensor partition strategy, we transform the problem into a multi-level image partitioning problem and propose the Multilevel Partition Algorithm (MPA). MPA achieves the same effect as non-penetrative partitioning and results in an efficient partitioning scheme.  





\section{Related Works}
Collaborative inference can effectively alleviate the computational and memory overhead of devices, reducing inference latency.\cite{hadidi2020toward,cao2019edge,hadidi2019robustly}. DeepThings \cite{zhao2018deepthings} integrates the early convolutional layers and distributes their computations across multiple devices in parallel.  MoDNN \cite{mao2017modnn} investigates the problem of additional communication overhead among IoT devices caused by image partition along multiple dimensions. CoEdge \cite{zeng2020coedge} builds on the application scenarios of MoDNN, introducing an improvement by designing a specialized hybrid parallelism strategy for deploying DNN models on IoT devices. However, both CoEdge and MoDNN employ a penetrative partitioning method for input images, indicating significant optimization potential for distributed inference in such scenarios.

In this paper, we propose NPTP, a non-penetrative partitioning approach for the original image as DNN input. NPTP can effectively reduce communication overhead between devices and decrease inference latency while maintaining the same device memory footprint.

\section{Problem Formulation and Algorithm}

\subsection{Computational Overhead Formulation}
The theoretical computation volume of the $i$-th image block at the $l$-th layer of the CNN can be denoted as Equation (1).
\begin{equation}
\label{deqn_ex1}
T^c_{l_i} = c^{in}_{l} \cdot c^{out}_{l} \cdot w^{k}_{l} \cdot h^{k}_{l} \cdot h^{out}_{l_i} \cdot w^{out}_{l_i}
\end{equation}
where $c^{in}_{l}$ and $c^{out}_{l}$ represent the number of input and output channels of the $l$-th convolutional operator in CNN model, respectively. $w^{k}_{l}$ and $h^{k}_{l}$ represent the kernel width and height of the $l$-th convolutional operator, respectively. $h^{out}_{l_i}$ and $w^{out}_{l_i}$ denote the height and width of the output feature map obtained after the $i$-th image block is processed by the $l$-th convolutional operator.

For $h^{out}_{l_i}$ and $w^{out}_{l_i}$ in Equation (1), we can compute them based on the properties of the convolution operation using Equation (2).
\begin{equation}
\begin{split}
\label{deqn_ex1}
h^{out}_{l_i} = \frac{h^{in}_{l_i} - h^{k}_{l} + 2p}{s_l}  + 1 \\
w^{out}_{l_i} = \frac{w^{in}_{l_i} - w^{k}_{l} + 2p}{s_l}  + 1
\end{split}
\end{equation}
where $h^{in}_{l_i}$ and $w^{in}_{l_i}$ are the height and width of the input feature map for the $i$-th image block at the $l$-th layer, $p$ is the padding size, and $s_l$ is the stride.

Thus, the total computational overhead for the $i$-th image block on the device $i$ can be denoted as Equation (3).
\begin{equation}
\label{deqn_ex1}
T^c_i = \sum_{l=1}^{N} \frac{T^c_{l_i}}{f_i}, i \in \mathcal{N}, l \in \mathcal{L} 
\end{equation}
where $f_i$ represents the clock frequency of IoT device $i$.

\subsection{Communication Overhead Formulation}

Equation (4) presents the communication volume, consisting of the number of rows and columns shared along the height and width dimensions of the image partition boundaries from adjacent devices.
\begin{equation}
\label{deqn_ex1}
T^g_{l_i} = P_{h_{l_i}} + P_{w_{l_i}}
\end{equation}
where $P_{h_{l_i}}$ and  $P_{w_{l_i}}$ represent the number of rows and columns, respectively.

Specifically, we can use the parameters of the convolutional operators to calculate $P_{h_{l_i}}$ and $P_{w_{l_i}}$ as Equation (5).
\begin{equation}
\begin{split}
\label{deqn_ex1}
P_{h_{l_i}} = s_{l} \cdot (h^{out}_{l_i}-1)-h^{in}_{l_i}+ s_{l} \\
P_{w_{l_i}} = s_{l} \cdot (w^{out}_{l_i}-1)-w^{in}_{l_i}+ s_{l}
\end{split}
\end{equation}
where $s_{l}$ denotes the stride of the $l$-th convolutional operator, $h^{in}_{l_i}$ and $w^{in}_{l_i}$ represent the height and weight of the input feature map of the $l$-th convolutional operator, respectively.

Thus, the total communication overhead for the $i$-th image block on the device can be denoted as Equation (6).
\begin{equation}
\label{deqn_ex1}
T^g_i = \sum_{l=1}^{N} \frac{T^g_{l_i}}{b_i}, i \in \mathcal{N}, l \in \mathcal{L}
\end{equation}
where $b_i$ represents the communication bandwidth of IoT device $i$.

\subsection{Inference Latency Formulation}

There are some numerical constraints on the partition sizes:
Equation (7) imposes the size restriction with the height and weight partition dimensions.
\begin{equation}
\begin{aligned}
\sum h_i & = H \\
\sum w_i & = W,\; i \in \mathcal{N}
\end{aligned}
\end{equation}
where W and H denote the size of the height and width of the original image. In other words, the combined heights or widths of the sub-images after partitioning are equal to those of the original input image.

Equation (8) imposes that the maximum peak memory usage during the inference process must be constrained by the device's memory capacity.

\begin{equation}
\begin{aligned}
\max (w_{l_i}+a_{l_i}) & \leq r_i, i \in \mathcal{N}, l \in \mathcal{L}
\end{aligned}
\end{equation}
where $w_{l_i}$ represents the memory consumption of the weights in the $l$-th layer of the CNN model, while $a_{l_i}$ represents the memory consumption of the activation values of image block $i$ at the output of the $l$-th layer.

Since the output of each IoT device must be aggregated before being fed into the classification stage to complete the subsequent inference, the total inference latency for the feature extraction phase of the model depends on the IoT device with the maximum latency in inference. Hence, we can formulate the distributed inference optimization as the following problem:
\begin{equation}
\begin{aligned}
\mathcal{P}1: & \quad \min_{a_i \in \eta} \, \max_{i \in \mathcal{N}} (T^{c}_{i} + T^{g}_{i}) \\
& \quad \text{s.t.} (7), (8).
\end{aligned}
\end{equation}

\subsection{Partitioning Algorithm Design}

\begin{figure}[!t]
\centering
\includegraphics[width=3.3in]{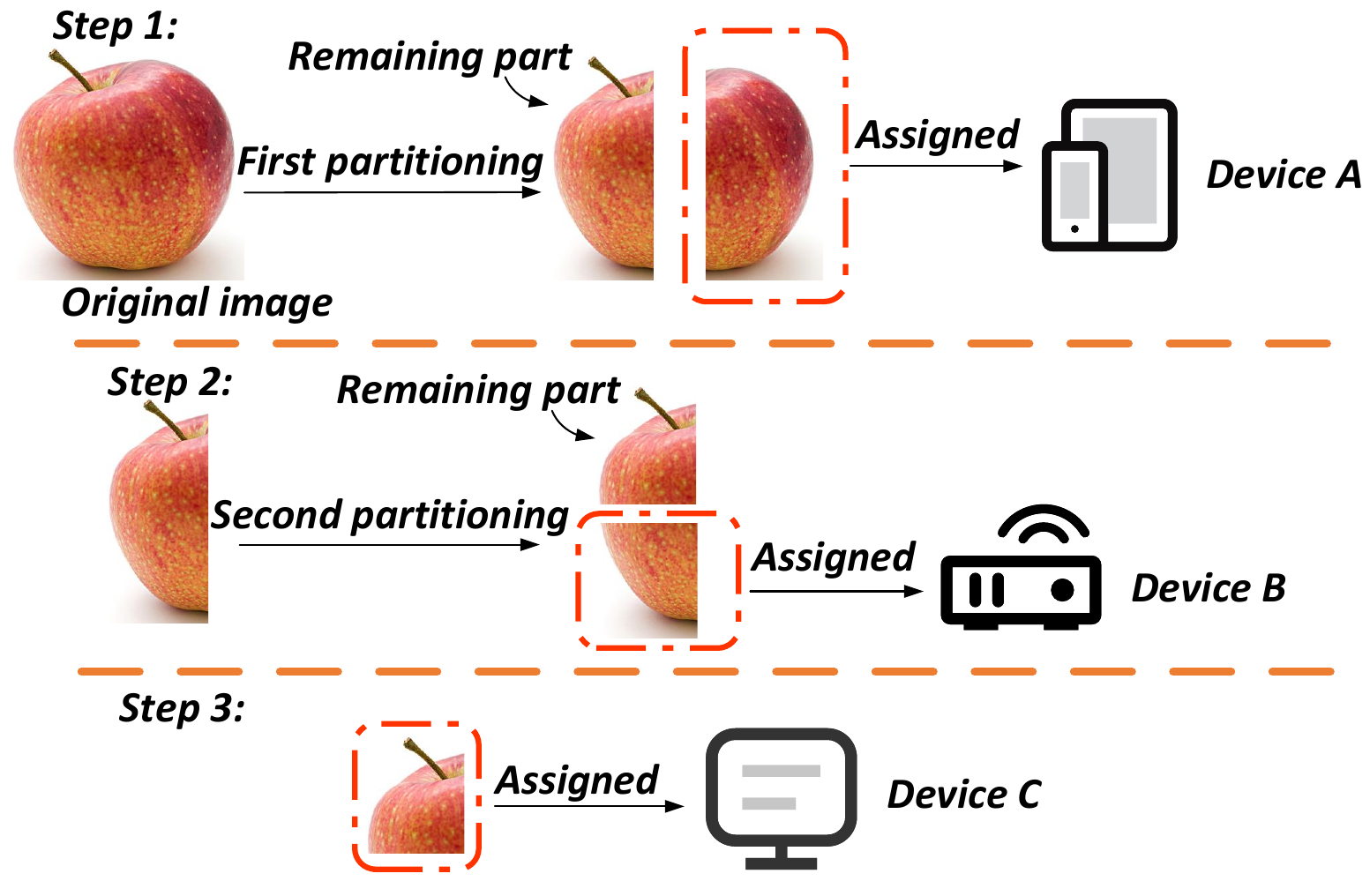}
\caption {Workflow overview of Multilevel Partitioning Algorithm (MPA). }
\label{fig_1}
\end{figure}


As the size of the partitioned image increases, the search space of $\mathcal{P}1$ becomes enormous. To improve search efficiency, we designed a heuristic policy Multilevel Partitioning Algorithm (MPA). This algorithm achieves the effect of non-penetrating partitioning equivalently by partitioning the original image multiple times. 

\begin{algorithm}[t!]
\caption{Multilevel Partitioning Algorithm (MPA)}
\label{alg:MPA}
\begin{algorithmic}[1]
\REQUIRE \quad \\
Original image size: $\mathcal{I}$ $(H, W)$ \\
CNN layers: $\mathcal{L} = [1, 2, \ldots, L]$ \\
Available Devices set: $\mathcal{N} = [1, 2, \ldots, N]$ \\
Device resources: $f_i, r_i, b_i, \forall i \in \mathcal{N}$ \\
CNN configurations: $(c^{in}, c^{out}, w^k, h^k, s, p)_l, \forall l \in L$ \\
Maximum iterations: $M$ \\
\ENSURE \quad \\
Assigned workload proportions: $\eta_{best}$
\STATE Initialize partition scheme list: $\eta$
\STATE Initialize minimum inference delay: $\mathcal{T}_{min} = \infty$
\STATE Initialize reward value: $\mathcal{E}$ = 1

\FOR{$i = 1$ to $M$}
    \STATE Initialize remaining image: $\mathcal{R} \leftarrow \mathcal{I}$
    

    \FOR{each device $N$ in $\mathcal{N}$}
            \STATE $l \leftarrow$ \textbf{select\_location}($\mathcal{R},\mathcal{E}$)
            \STATE $\mathcal{P} \leftarrow$ \textbf{partition\_image}($\mathcal{R}$, $l$)
            \STATE $\mathcal{R} \leftarrow$ \textbf{update\_remaining\_image}($\mathcal{R}$, $\mathcal{P}$)
            \STATE Add to current partition scheme: $\eta \leftarrow \eta_i$
    \ENDFOR   
 
    \STATE $\mathcal{T} \leftarrow$
    \textbf{evaluator}($\eta$, $\mathcal{L}$, $\mathcal{N}$)
    
    
    \IF{$\mathcal{T} < \mathcal{T}_{min}$}
        \STATE Update minimum delay: $\mathcal{T}_{min} \leftarrow \mathcal{T}$
        \STATE Record best partition scheme: $\eta_{best} \leftarrow \eta$
        \STATE Update reward value: $\mathcal{E} = 1$
    \ELSE
        \STATE Update reward value: $\mathcal{E} = -1$
    \ENDIF
\ENDFOR
\RETURN $\eta_{best}$

\end{algorithmic}
\end{algorithm}

As shown in Fig. 3, step 1 selects either the height or width dimension of the original image for partitioning to obtain sub-image 1, which is then assigned to device A as input. In step 2, a portion of the remaining image is partitioned to form sub-image 2, which is assigned to device B. This process is repeated until the entire image is partitioned and assigned. This method equivalently implements a non-penetrating partition of the original image. The generated partition scheme is input to the evaluation function, where the corresponding inference delay is calculated according to Equation (3) and Equation (6). From the second generation of the partition scheme onwards, each generated partition scheme is rewarded or penalized by comparing it with the last obtained scheme. After completing the specified number of iterations for the predefined round, the partition and assignment scheme with the highest reward is selected as the final solution. The detailed computational process for a single round is presented in Algorithm 1.

\section{Experiments and results}

We implemented a prototype of NPTP using three NVIDIA graphics cards to simulate IoT devices. To achieve a heterogeneous model scenario, we limited the memory capacity of each GPU using the pytorch API as $torch.cuda.set$. We implemented the VGG13, VGG16, and VGG19 networks \cite{simonyan2014very} using PyTorch in our prototype. These classic neural network models have different numbers of convolutional operators in their feature extraction stage. 



\textbf{Impacts of bandwidths} We evaluate the inference latency of these network models under both the CoEdge partitioning scheme and the NPTP partitioning scheme in a scenario where the bandwidth between devices ranges from 0.1 MB/s to 1 MB/s, as shown in Fig. 4. In comparison to the CoEdge partitioning scheme, NPTP achieved speedup by factors of 1.22-1.31$\times$, 1.32-1.43$\times$, 1.37-1.52$\times$, and 1.45-1.58$\times$ for VGG11, VGG13, VGG16, and VGG19, respectively. We observe that the improvement of the NPTP scheme is more significant on VGG19 than on the other three models. This is because VGG19 has more convolutional layers, which leads to a reduction in data-sharing overhead at partition boundaries during inference.

\begin{figure}[!t]
\subfloat{
		\includegraphics[scale=0.25]{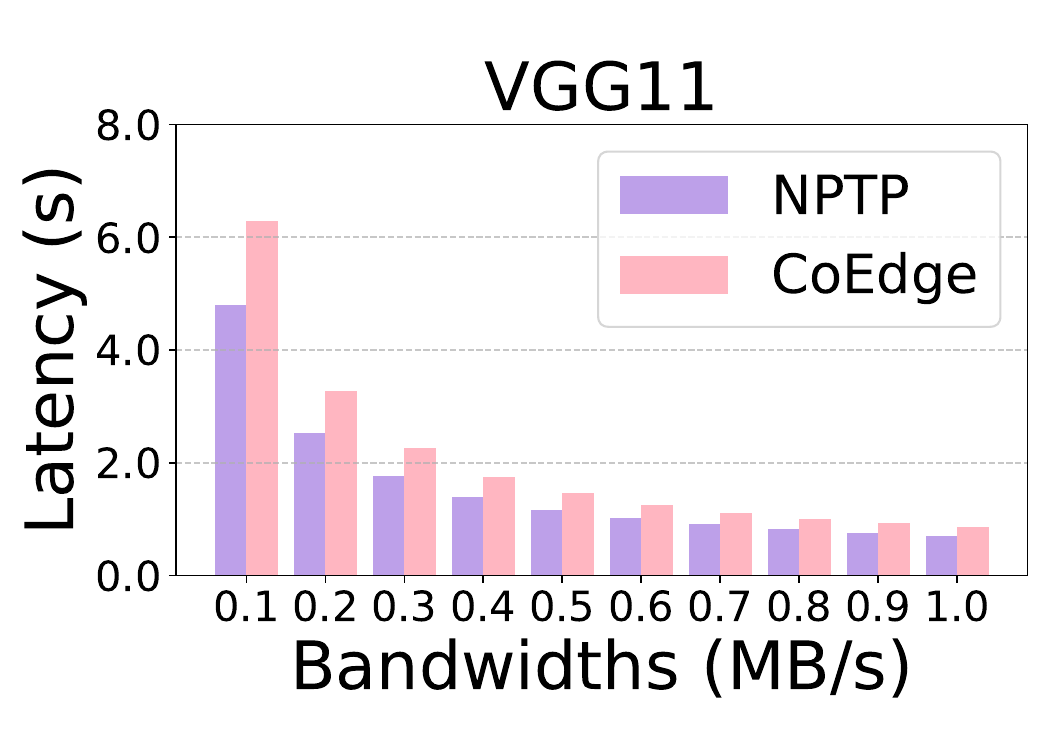}}
\subfloat{
		\includegraphics[scale=0.25]{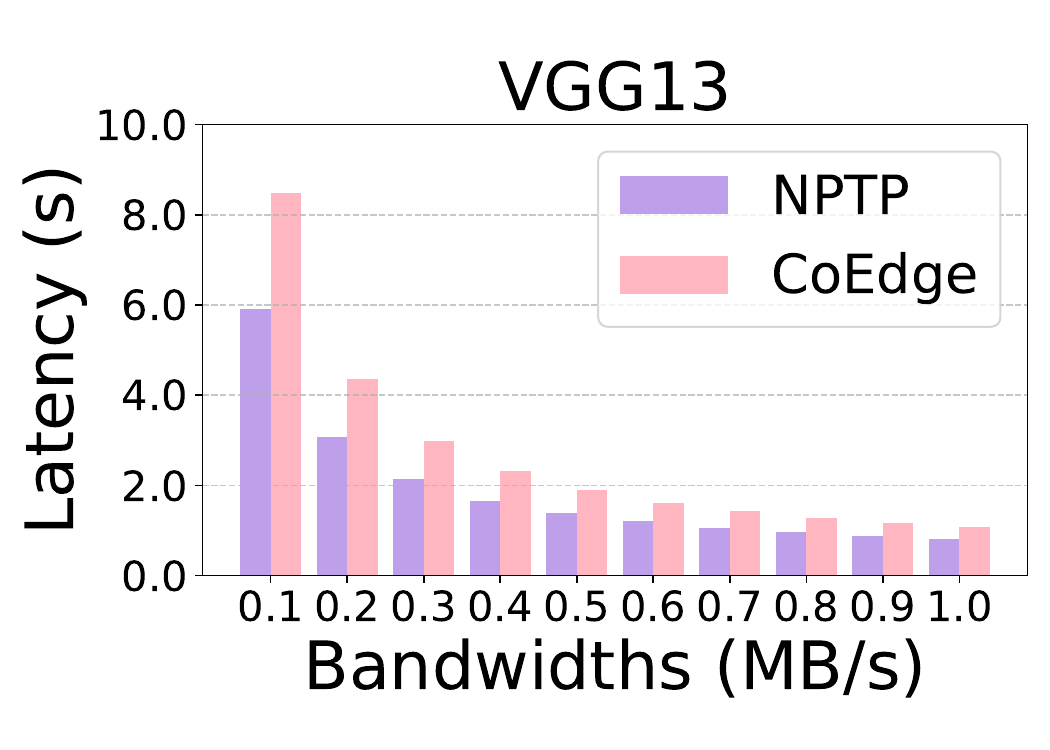}}
\\
\subfloat{
		\includegraphics[scale=0.25]{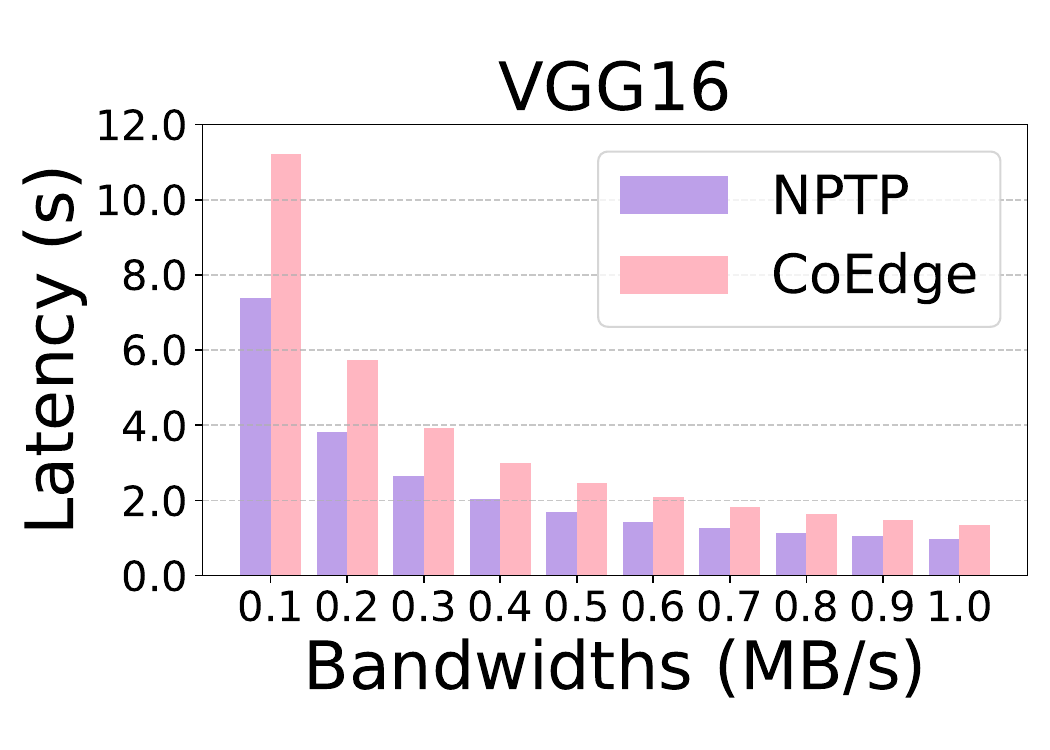}}
\subfloat{
		\includegraphics[scale=0.25]{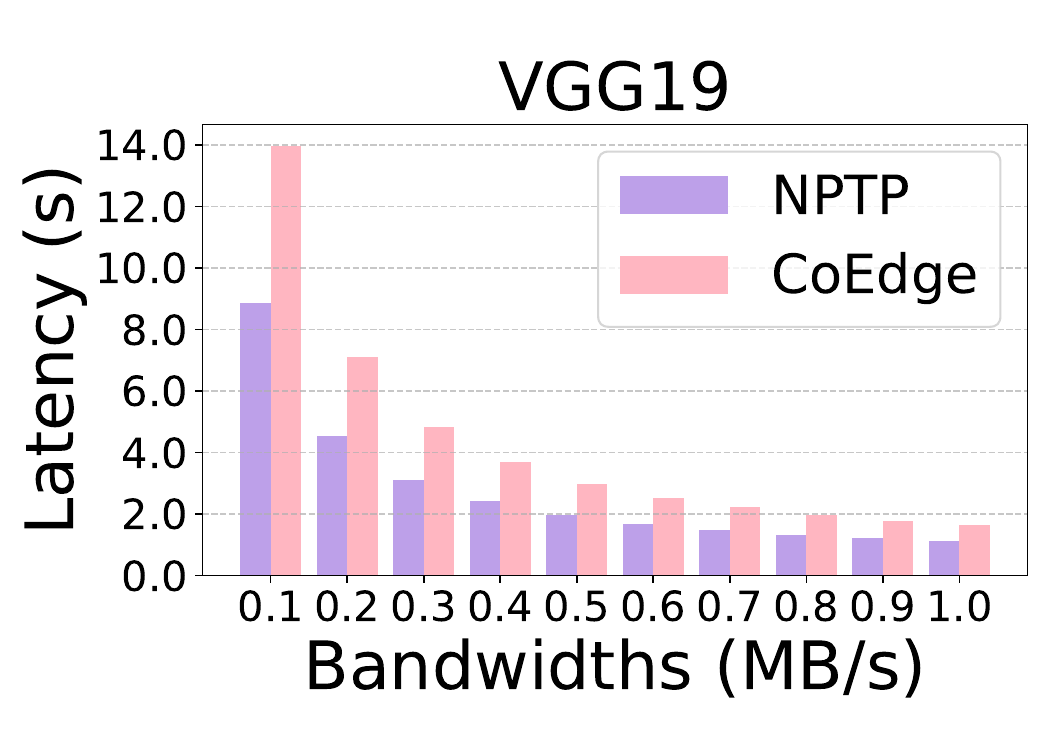}}
\caption{Inference latency of NPTP and CoEdge partitioning schemes under different device communication bandwidths.}
\label{fig_6}
\end{figure} 

\begin{figure}
\subfloat{
		\includegraphics[scale=0.125]{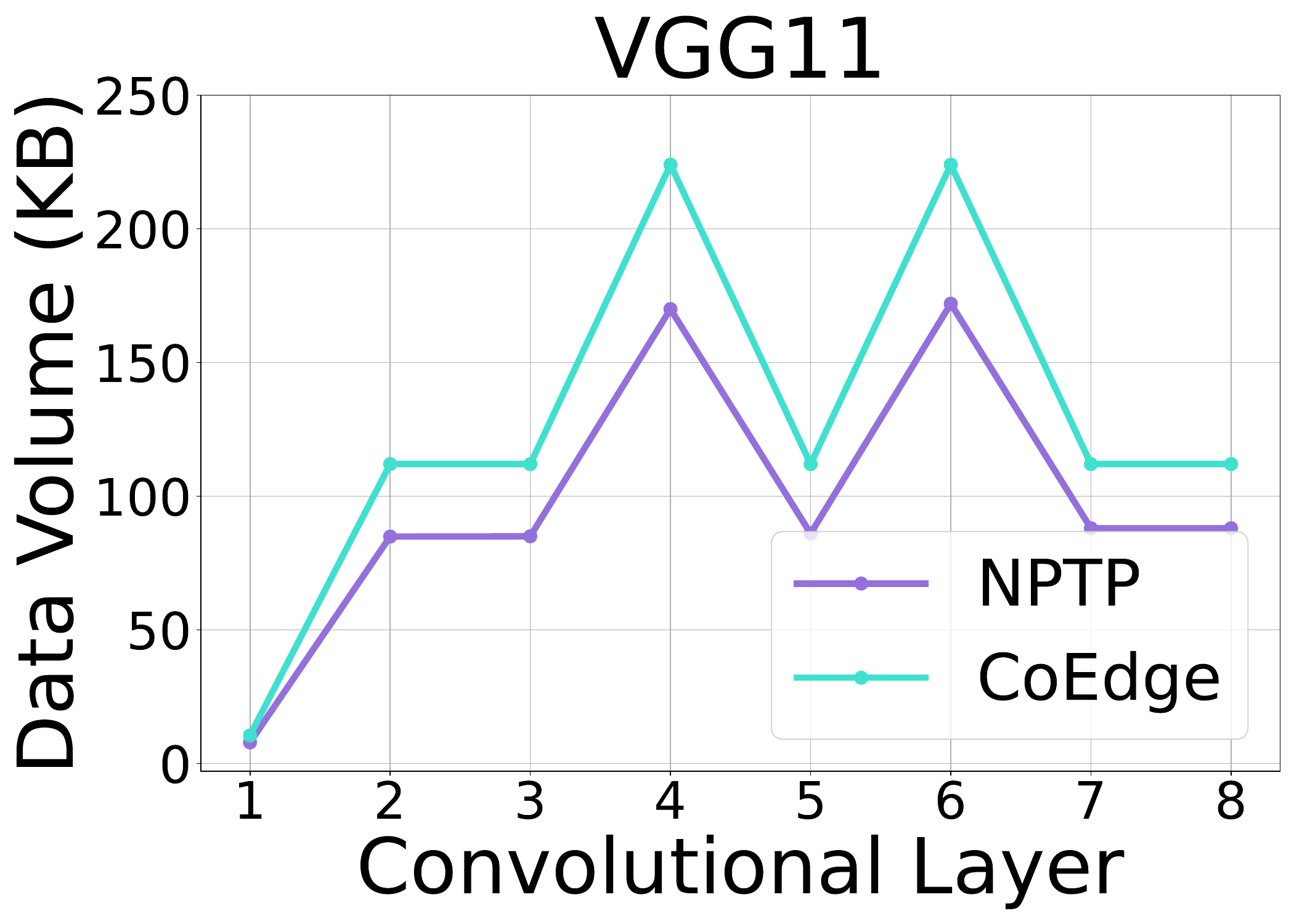}}
\subfloat{
		\includegraphics[scale=0.125]{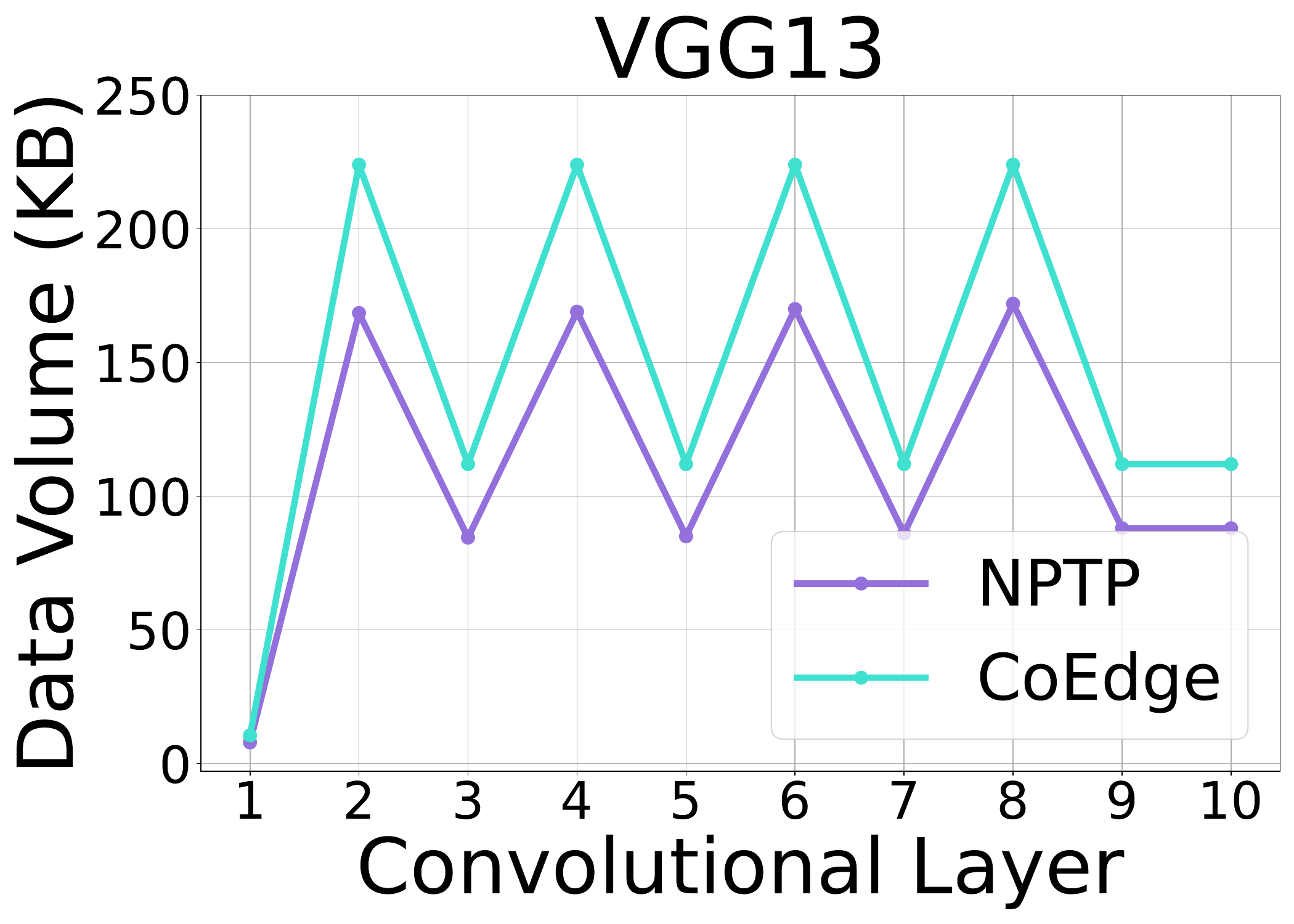}}
\\

\subfloat{
		\includegraphics[scale=0.125]{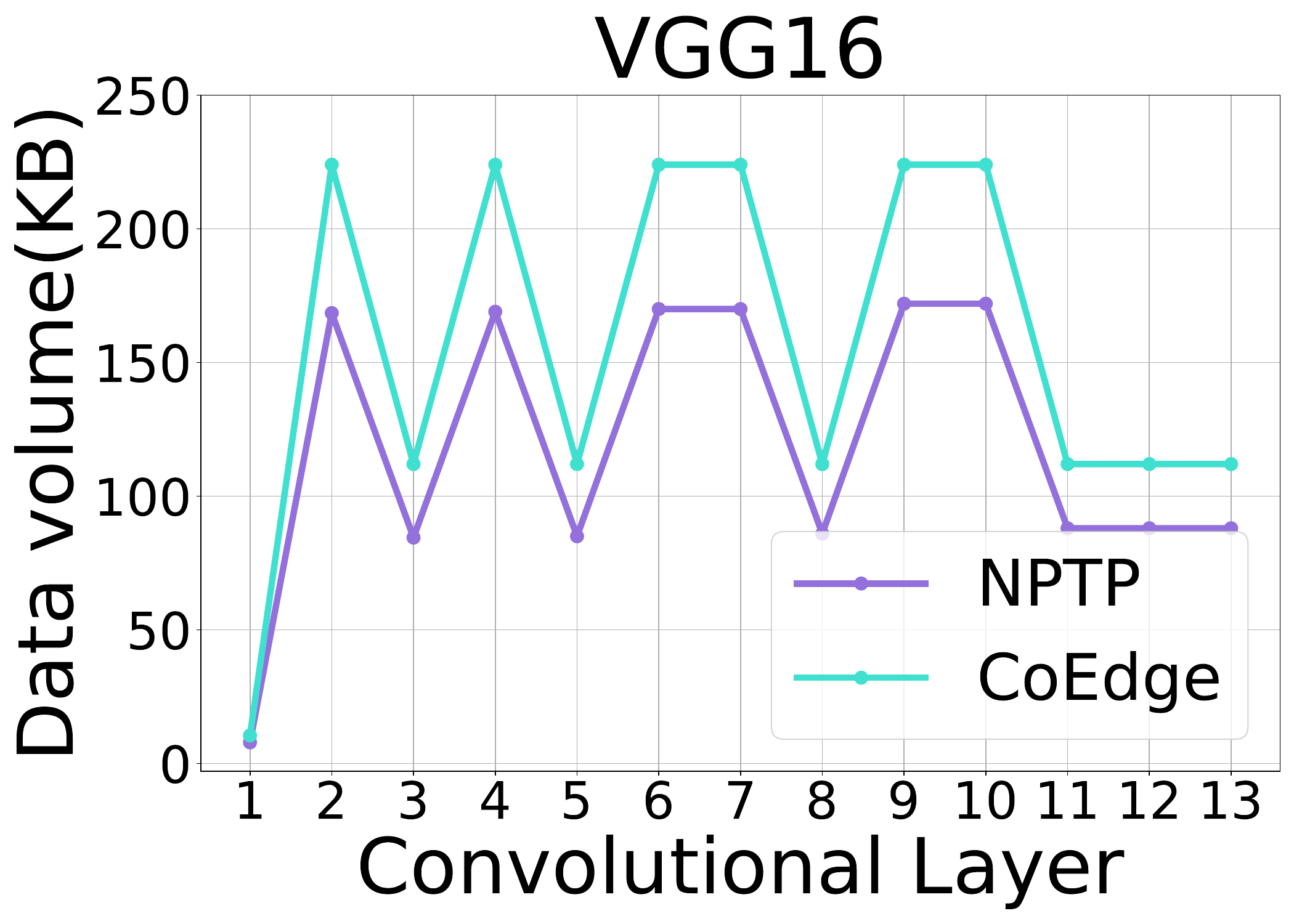}}
\subfloat{
		\includegraphics[scale=0.125]{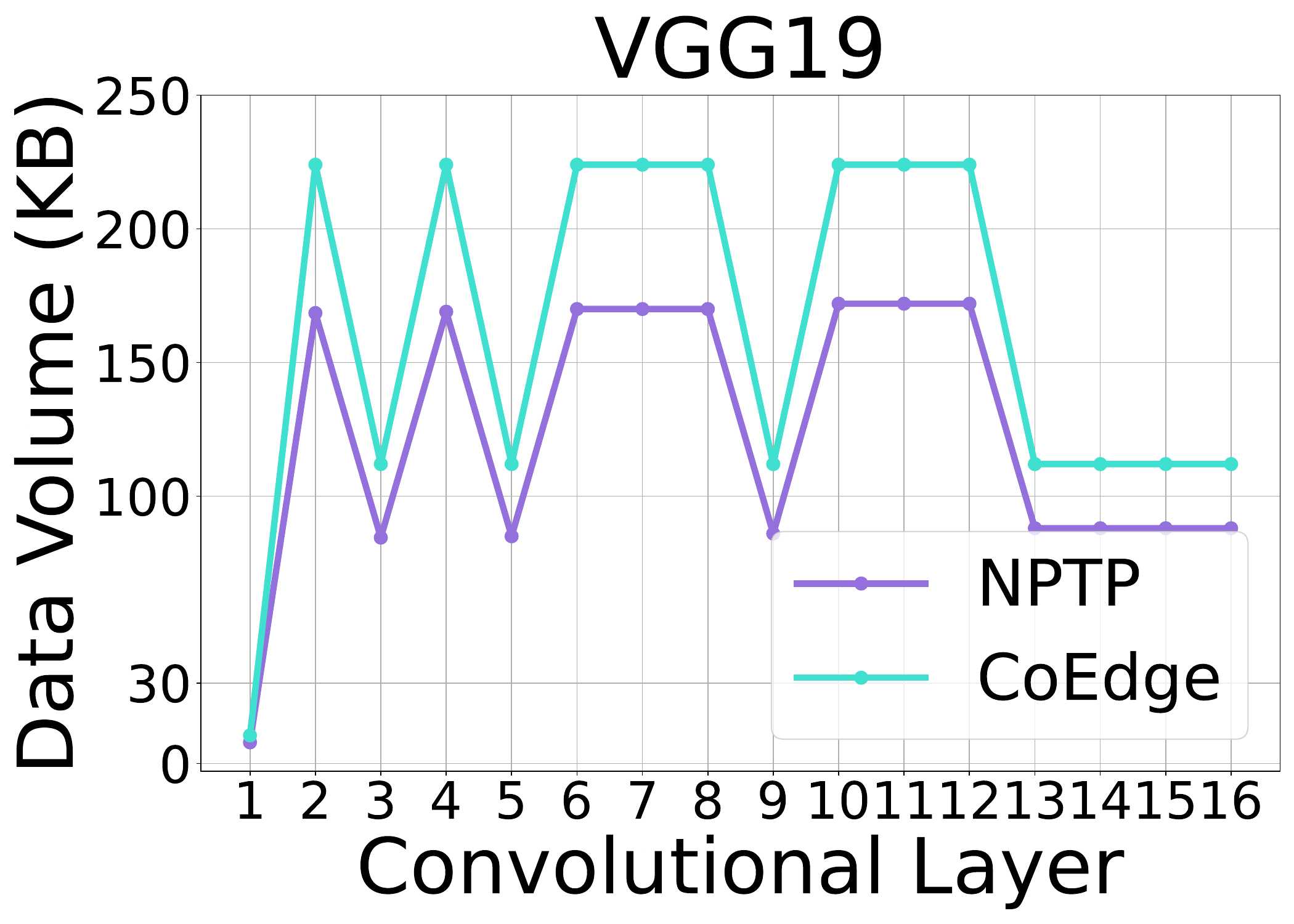}}
\caption{Communication data volume of NPTP and CoEdge partitioning schemes under different models.}
\label{fig_6}
\end{figure}

In order to conduct a quantitative analysis of the communication savings during the inference stage, the inter-device communication for each layer's operators. As shown in Fig. 5, NPTP can reduce up to 1.32$\times$ communication volume compared to CoEdge. During the model inference process, the NPTP scheme consistently has lower inter-device communication than CoEdge during the execution of each operator. The trend of communication volume is approximately the same for both partitioning schemes. This is because when the partitioning method of the input image is determined, each partition inputs the same model and performs the same computational process.

\textbf{Impacts of image size} Since the NPTP scheme is proposed to effectively address the issue of peak memory usage exceeding device capacity during inference on large images, it is intuitive to investigate how the selection of large image sizes affects its performance. In this experiment, we employed the VGG13 and VGG16 models, distributing them across three devices. We evaluated their performance with varying image sizes: 224 × 224, 512 × 512, 1024 × 1024, 2048 × 2048, and 4096 × 4096. As shown in Fig. 6, NPTP consistently outperformed CoEdge across all image sizes, achieving 1.44-1.68× and 1.47-1.64× inference speedup on the VGG13 and VGG16 networks, respectively.

\begin{figure}[!h]
\subfloat{\includegraphics[width=1.78in]{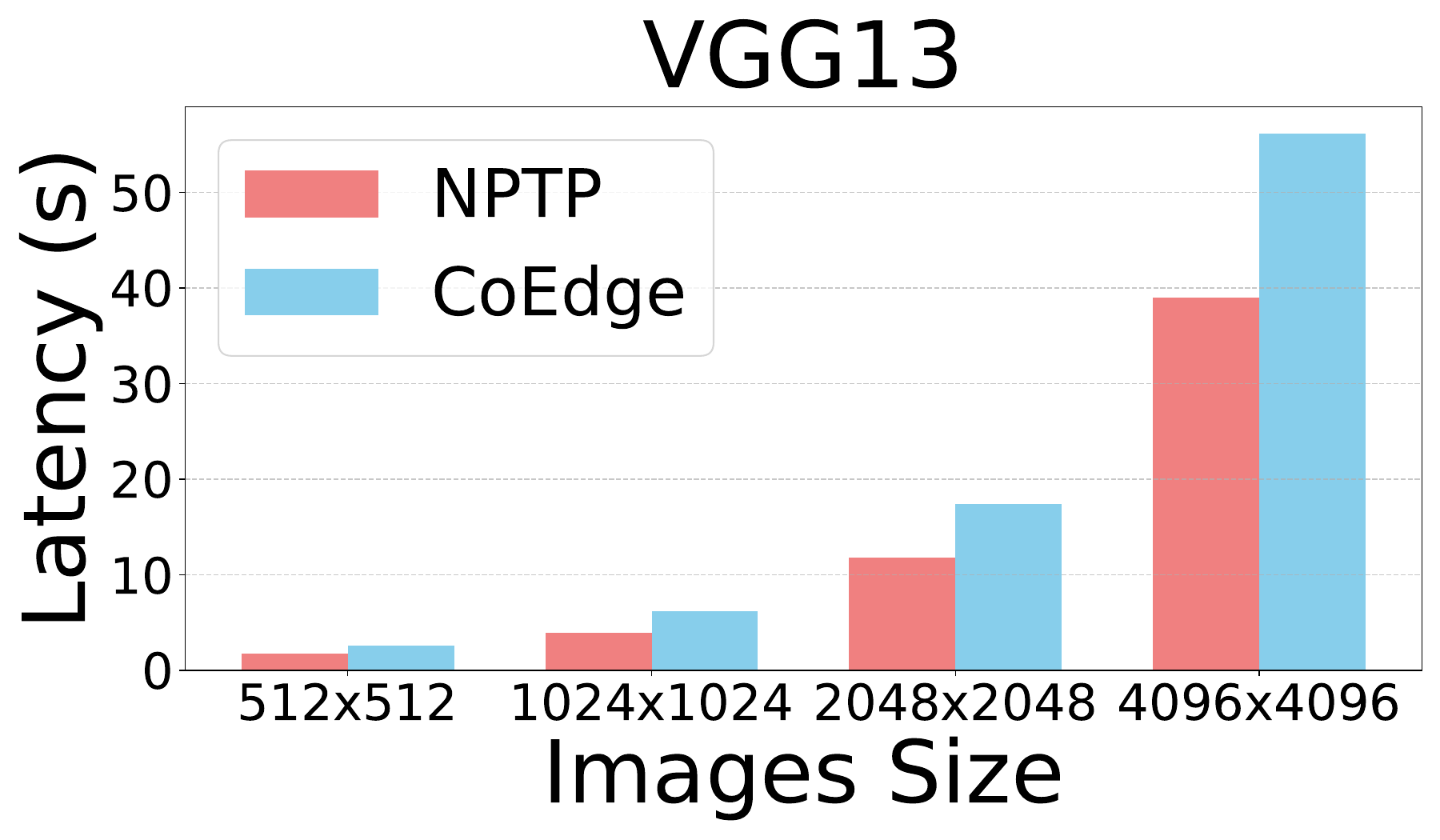}%
\label{Convolutional computation process for the complete image}}
\subfloat{\includegraphics[width=1.78in]{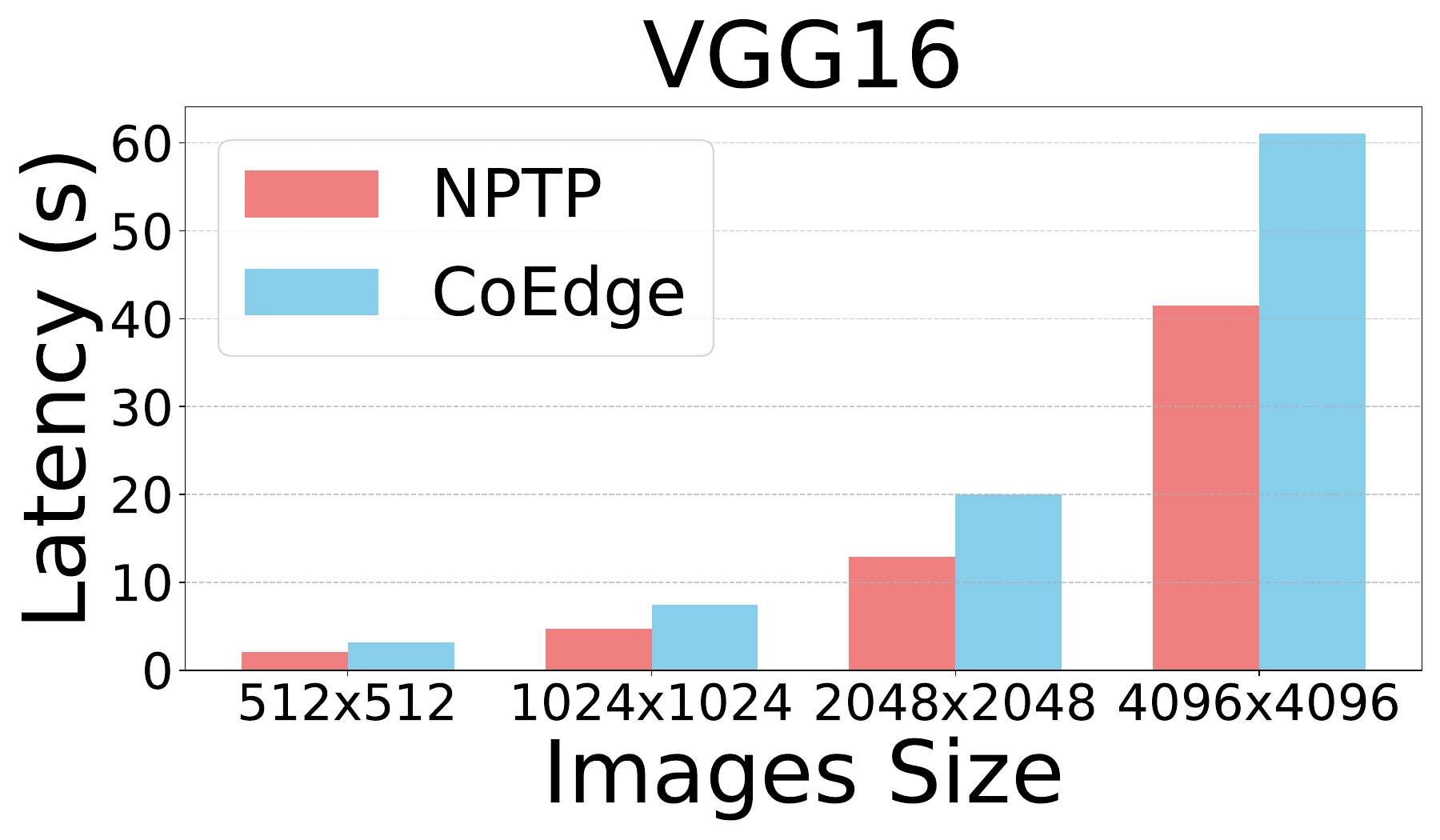}%
\label{fig_second_case}}
\caption{NPTP’s performance using different image sizes as model input.}
\label{fig_sim}
\end{figure}



\section{Conclusions}

In this paper, we propose NPTP, a novel collaborative inference scheme that achieves significant speedup by reducing the data sharing overhead caused by partitioning boundaries for convolutional operations. A low-complexity heuristic algorithm, MPA, is designed to find the optimal NPTP partitioning schemes. MPA effectively achieves non-penetrating partitioning by partitioning the original image multiple times and introducing penalty and reward terms to evaluate the partitioning schemes. Our empirical analysis demonstrates that it achieves a 1.44–1.68× inference speedup compared to CoEdge, a SOTA collaborative inference system.


\section*{Acknowledgement}
This study is supported by the National Key R\&D Program of China (Grant No. 2022YFB4501600). 

\end{document}